# ZeroMat: Solving Cold-start Problem of Recommender System with No Input Data

Hao Wang
Ratidar.com
Beijing, China
haow85@live.com

***Abstract*—Recommender system is an applicable technique in most E-commerce commercial product technical designs. However, nearly all recommender system faces a challenge called the cold-start problem. The problem is so notorious that almost every industrial practitioner needs to resolve this issue when building recommender systems. Most cold-start problem solvers need some kind of data input as the starter of the system. On the other hand, many real-world applications place popular items or random items as recommendation results. In this paper, we propose a new technique called ZeroMat that requries no input data at all and predicts the user item rating data that is competitive in Mean Absolute Error and fairness metric compared with the classic matrix factorization with affluent data, and much better performance than random placement.**

## I. INTRODUCTION

Recommender systems are crucial in user acquisition and monetization of web products. More than 30% of products sold on Amazon.com are due to recommendation. It would cost much more money by scale to entice customers for the same amount of purchases via other venues such as conventional ads or Google Adwords. Therefore, most E-commerce companies have spent a tremendous amount of investment in recommender system research and development.

Every recommender system receives new users and new items. However, these new data records have no history in the past that facilitate the recommendation. This problem is known as the cold start problem, and is faced by every recommender system builder in the industry. Some products select random items or popular items and recommend them to new users, or use content-based recommendation in certain contexts such as news recommendation to resolve the cold-start problem for new items. Other more rigorous methods utilize algorithms to alleviate the cold-start problem, but they more or less rely on some data inputs as the starter to solve this issue.

In this paper, we propose a new method called ZeroMat to resolve this problem. Our method is inspired by Probabilistic Matrix Factorization. By modeling the user item rating distribution using Zipf's Law, we acquire a new approach that resolves the cold start problem with no need of input data. In the Experiment Section, we illustrate that our method is mostly competitive with the classic matrix factorization with affluent input data.

It must be surprising that a recommender system without input data can yield good predictions for user-rating-matrix, but the experimental result serves as very strong evidence to support the validity of our method. This might lead to the conclusion that although recommender system has developed for a couple of decades, we are really still not doing very well in accuracy and fairness metrics.

## II. RELATED WORK

Recommender system has a decade-long history of enhancing commercial values for business firms. Famous recommender system techniques such as collaborative filtering [1], matrix factorization [2][3][4], learning to rank[5][6], deep-learning based recommendation [7][8] have all been applied on large scales in companies throughout the world [9].

Recommender system faces notorious challenges such as Matthew Effect problem and Cold-start Problem. As a special case of the fairness problem, Matthew Effect has raised awareness among researchers such as Wang [10][11][12] in recent years, and fairness problem [13][14][15] in general is a hot research topic in many research conferences.

Unlike the emerging interest on the Matthew Effect problem, the cold-start problem is usually considered as being more challenging and unsolvable. Very few researchers have shed some insightful light on the topic [16][17][18], and most of them require some sort of data input as the starter to solve the issue.

## III. PROBLEM FORMULATION

Probabilistic Matrix Factorization problem is a framework equivalent to the classic matrix factorization. The classic matrix factorization is formulated as follows:

$$RMSE = \sum_{i=1}^{m} \sum_{j=1}^{n} \left( u_i \bullet v_j - R_{ij} \right)^2$$

The Probabilistic Matrix Factorization [19] formulates the problem as a maximum likelihood problem.

$$P(R \mid U, V, \sigma_U, \sigma_V) \sim N(R \mid U, V, \sigma_U, \sigma_V)$$

, and we are looking for the solution that maximizes the following maximum posterior:

$$P(U,V \mid R,\sigma_U,\sigma_V)$$

By taking the natural log of the maximum posterior and simple mathmatical calculation, it can be shown the Probabilistic Matrix Factorization paradigm is equivalent to the classic matrix factorization approach with regularization.

We formulate ZeroMat by Zipf's Law as follows:

$$\frac{R_{i,j}}{R_{\max}} \sim \frac{U_i \bullet V_j}{\max(U_i \bullet V_j)}$$

We omit maximum value of the dot product for now and represent the ratio of rating scores as R and formulate the maximum posterior of R below:

$$P(R \mid U,V,\sigma_U,\sigma_V) \sim \prod_{i=1}^{N}\prod_{j=1}^{M}\left(U_i^T \bullet V_j\right)$$

We further dictate that U and V follows the zero-mean normal distribution as specified in the Probabilistic Matrix Factorization framework. We acquire the following posterior formula:

$$P(U,V \mid R,\sigma_U,\sigma_V) = \frac{P(R \mid U,V,\sigma_U,\sigma_V) \times P\left(U,V \mid \sigma_U,\sigma_V\right)}{P(R,\sigma_U,\sigma_V)}$$

We omit the joint distribution of hyper-parameters and acquire the following formula:

$$P(U,V \mid R,\sigma_U,\sigma_V) \sim P(R \mid U,V,\sigma_U,\sigma_V) \times P(U,V \mid \sigma_U,\sigma_V)$$

Once again, we assume the probability distribution of U and V are zero-mean normal distributions. We plug in the probability distributions into the posterior formula of U and V:

$$P(U,V \mid R,\sigma_U,\sigma_V) = \prod_{i=1}^{N}\prod_{j=1}^{M}\left(U_i^T \bullet V_j\right) \times \prod_{i}^{N} e^{\frac{-U_i^T \bullet U_i}{2\sigma_U^2}} \times \prod_{i}^{M} e^{\frac{-V_i^T \bullet V_i}{2\sigma_V^2}}$$

After taking the natural log of the posterior of U and V, we obtain the loss function L for optimization:

$$L = \sum_{i=1}^{N}\sum_{j=1}^{M} \ln(U_i^T \bullet V_j) - \frac{1}{2\sigma_U^2} \times \sum_{i=1}^{N} U_i^T \bullet U_i - \frac{1}{2\sigma_V^2} \times \sum_{i=1}^{M} V_i^T \bullet V_i$$

we can solve for the optimal values of U and V that maximize L, i.e.:

$$U_i = U_i + \eta \times \left(\frac{V_j}{U_i^T \bullet V_j} - 2 \times U_i\right)$$

$$V_j = V_j + \eta \times \left(\frac{U_i}{V_j^T \bullet U_i} - 2 \times V_j\right)$$

. Please notice the computation of U and V require no knowledge of the user rating values R.

We reconstruct the values of the user ratings using U and V by the following formula:

$$R_{i,j} = R_{\max} \times \frac{U_i \bullet V_j}{\max(U_i \bullet V_j)}$$

The estimated rating values can be used for cold start problem. The entire algorithmic procedure requires no knowledge of user rating matrix. The only thing we need to know is the maximum value of the ratings. This is usually a product design knowledge that we know beforehand. For example, MovieLens dataset has a rating scale ranging from 1 to 5, then we know $R_{\max}$ is 5.

In the next section, we compare our algorithm in technical accuracy metric and fairness metric against the random value placement heuristics and the classic matrix factorization solved by Stochastic Gradient Descent. The metrics that we use are Mean Absolute Error (MAE) for accuracy performance and the Degree of Matthew Effect (Wang [12]) for fairness evaluation.

## IV. Experiment

We use the small dataset of MovieLens that include 610 users and 9724 items to test ZeroMat against random placement and the classic matrix factorization solved by Stochastic Gradient Descent. We do a grid search on the gradient learning step to check the overall performance and the robustness of the method (Fig. 1 and Fig. 2) :

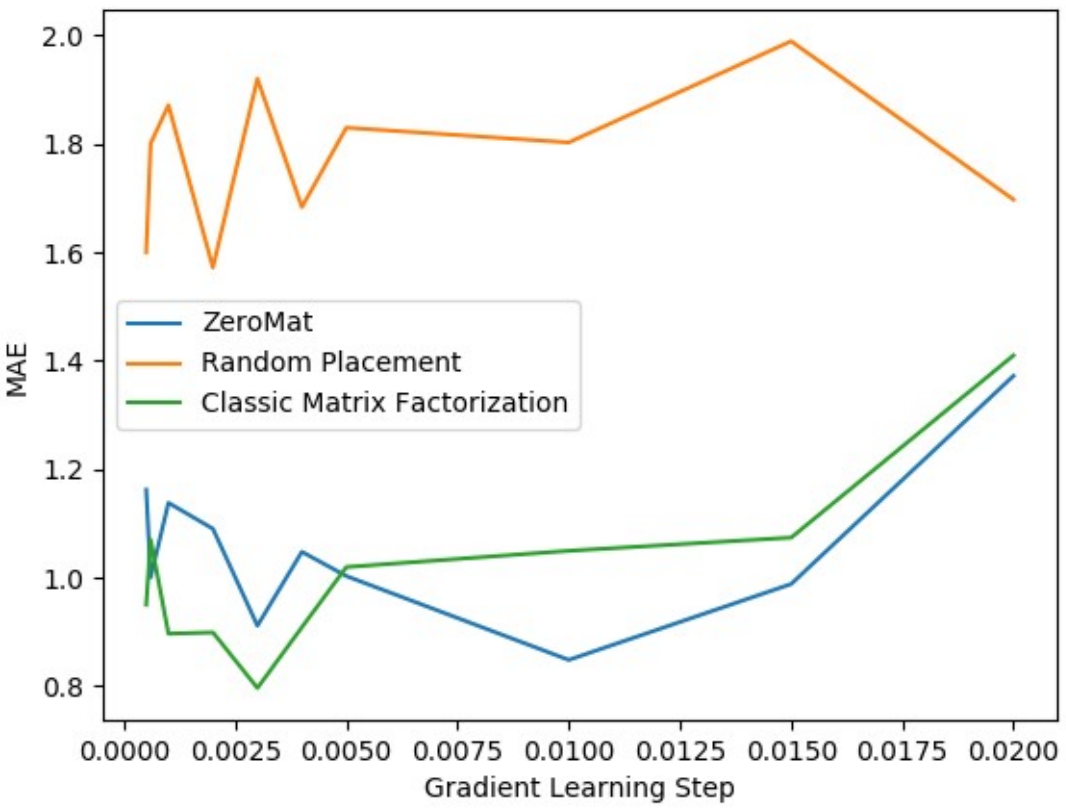

Fig.1 MAE comparison among ZeroMat, Random Placement and the Classic Matrix Factorization

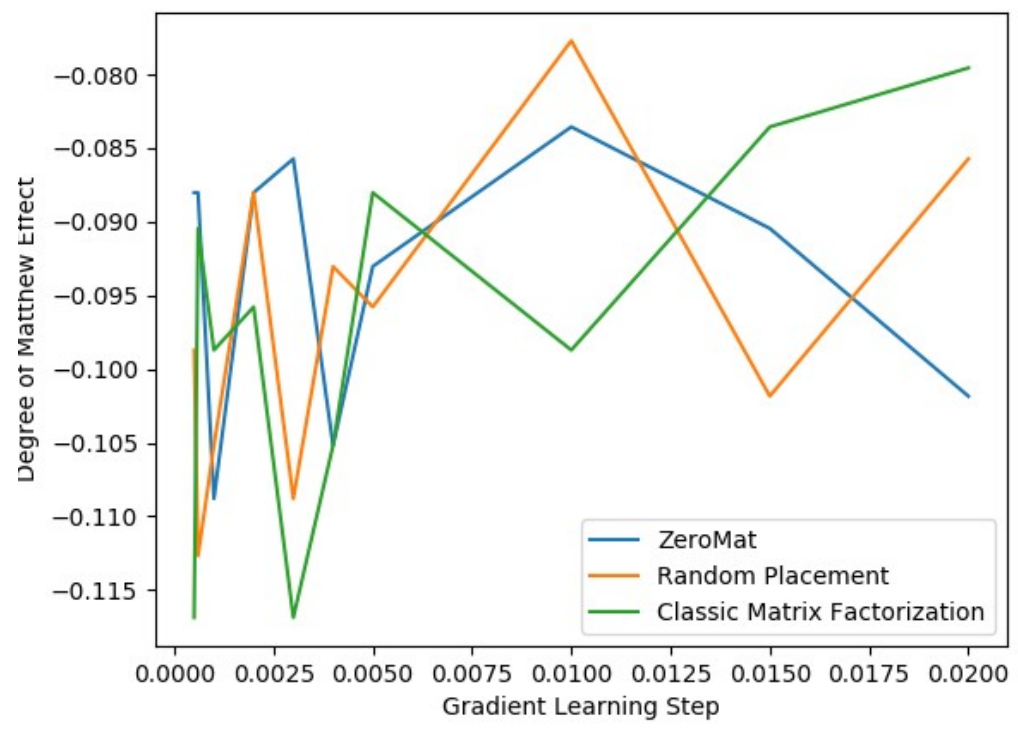

Fig.2 Degree of Matthew Effect comparison among ZeroMat, Random Placement and the Classic Matrix Factorization

The small scale dataset demonstrates that our algorithm is competitive with the classic matrix factorization approach. ZeroMat even outperforms the classic matrix factorization in MAE for a large number of parameters, and its worst performance is less than 30% worse than the classic matrix factorization. ZeroMat outperforms random placement by a large margin in MAE. As for fairness metric, the 3 methods are comparable with each other.

For large scale dataset (Fig. 3 and Fig. 4), we test our approach on MovieLens 1 million dataset that comprises 6040 users and 3706 items. The result is consistent with MovieLens small dataset (Fig. 1 and Fig. 2). However, due to randomness, in some cases, with competitve MAE score, ZeroMat outperforms the classic matrix factorization in fairness metric by large margin (Fig.5).

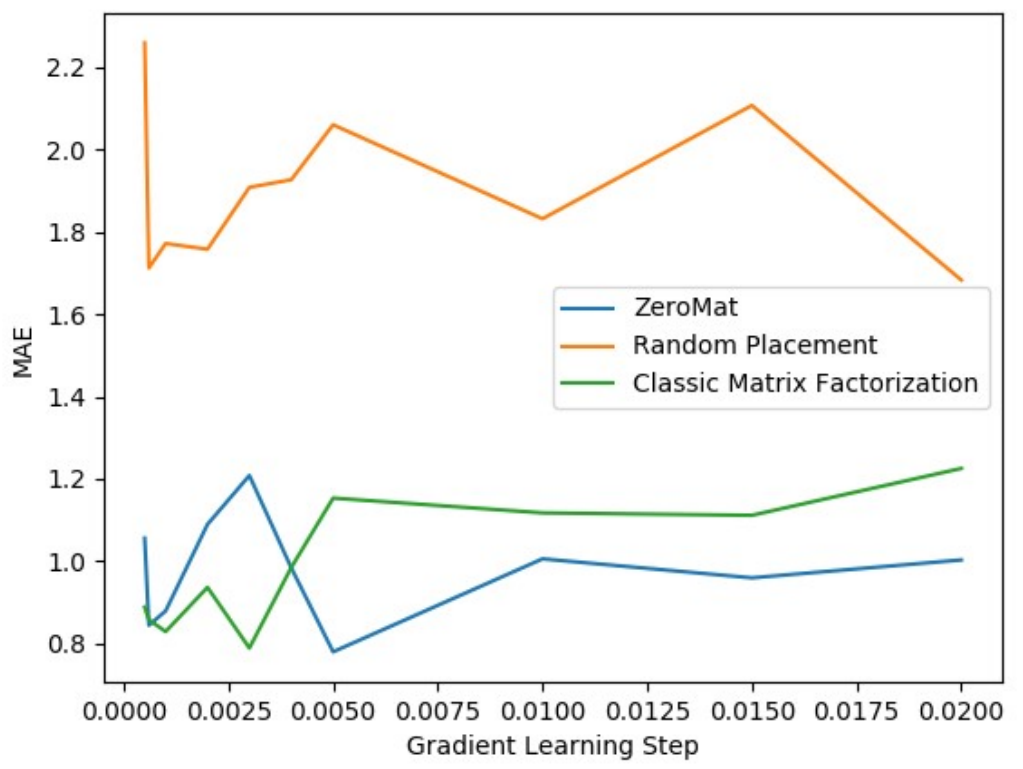

Fig.3 Comparison in MAE on MovieLens 1 Million Dataset

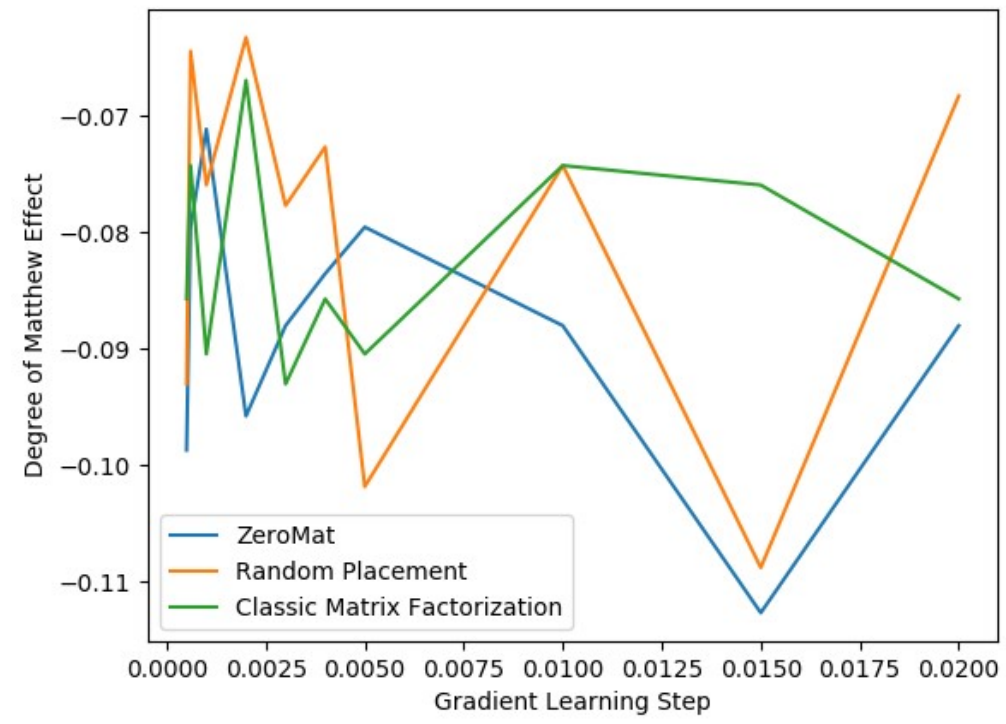

Fig.4 Comparison in fairness on MovieLens 1 Million Dataset

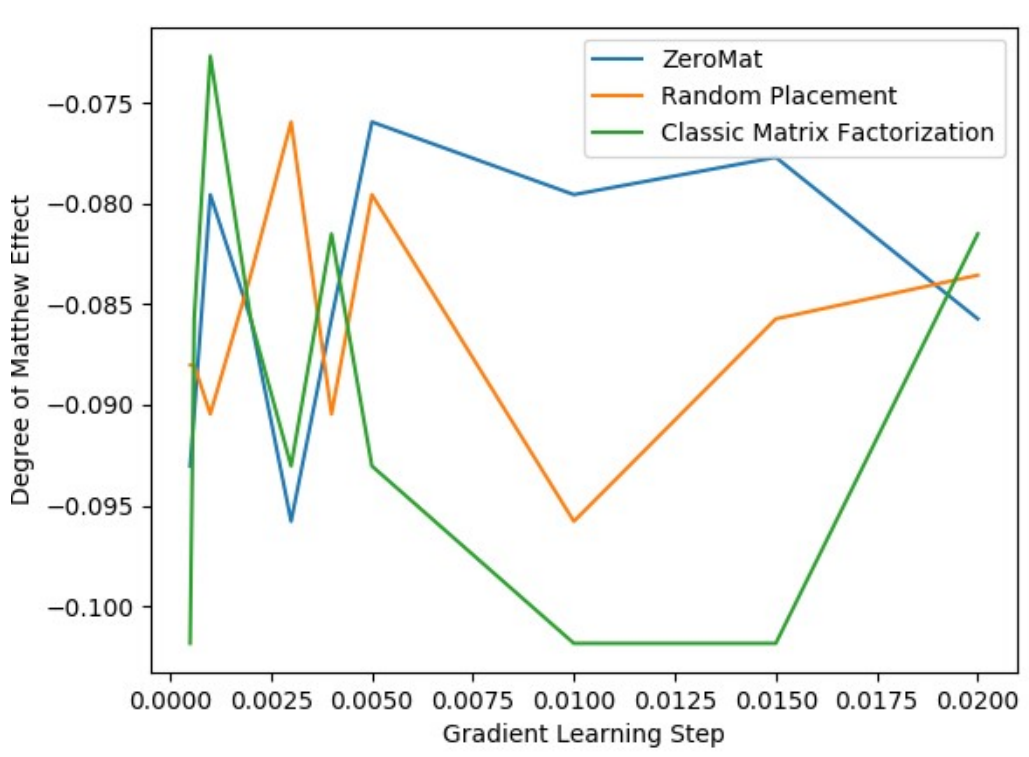

Fig.5 Comparsion in fairness metric on MovieLens 1 Million Dataset

## V. Conclusion

In this paper, we propose a new method called ZeroMat that requires no data input as the starter to solve the cold start problem. Our approach outperforms the random recommendation by a large margin. Even compared with the classic matrix factorization, our method is competitive in both MAE and fairness metric. The implication is two folds : 1. Our method is an ideal tool for the cold-start problem. 2. Matrix factorization techniques are far from being good enough.

In future work, we would like to explore in depth the reason why ZeroMat can outperform the classic matrix factorization technique in a large range of gradient step learning parameters.

## References

[1] S.Xiaoyuan and M.K.Taghi, A Survey of Collaborative Filtering Techniques, Advances in Artificial Intelligence, 2009

[2] T. Chen, W. Zhang, Q. Lu, K. Chen, Y. Yu. SVDFeature：A Toolkit for Feature-based Collaborative Filtering. The Journal of Machine Learning Research, 2012.

[3] Y. Koren. Factorization Meets the Neighborhood: a Multifaceted Collaborative Filtering Model. KDD, 2008.

[4] G. Takacs, D. Tikk. Alternating Least Squares for Personalized Ranking. The 6th ACM Conference on Recommender Systems, 2012.

[5] W. Pan, Z. Hao, C. Xu, et al. Adaptive Bayesian personalized ranking for heterogeneous implicit feedbacks. Knowledge-Based Systems, 2015.

[6] S. Yue , M. Larson , A. Karatzoglou, et al. CLiMF: Collaborative Less-Is-More Filtering, Proceedings of the Twenty-Third International Joint Conference on Artificial Intelligence , 2014.

[7] J.Hong, X.Dai,J.Zhang, S.Huang, J. Chen, Deep Matrix Factorization Models for Recommender Systems, Proceedings of the Twenty-Sixth International Joint Conferenceon on Artificial Intelligence, 2017.

[8] L.Chen , H.Shi. DexDeepFM: Ensemble Diversity Enhanced Extreme Deep Factorization Machine Model[J]. 2021

[9] T.Chen, J.Cai, H.Wang, D.Yu, Instant Expert Hunting: Building an Answerer Recommender System for a Large Scale Q&A Website", ACM Symposium on Applied Computing, 2014

[10] H. Wang, B. Ruan. MatRec: Matrix Factorization for Highly Skewed Dataset. The 3rd International Conference on Big Data Technologies, 2020.

[11] H. Wang, Z. Wang, W. Zhang. Quantitative Analysis of Matthew Effect and Sparsity Problem of Recommender Systems, IEEE International Conference on Cloud Computing and Big Data Analysis， 2018.

[12] H. Wang. Zipf Matrix Factorization: Matrix Factorization with Matthew Effect Reduction. The 4th International Conference on Artificial Intelligence and Big Data, 2021.

[13] D. Kiswanto , D. Nurjanah , and R. Rismala . Fairness Aware Regularization on a Learning-to-Rank Recommender System for Controlling Popularity Bias in E-Commerce Domain. International Conference on Information Technology Systems and Innovation (ICITSI) 2018

[14] R. Cañamares, P. Castells. Should I Follow the Crowd? A Probabilistic Analysis of the Effectiveness of Popularity in Recommender Systems. The 41st International ACM SIGIR Conference on Research & Development in Information Retrieval, 2018.

[15] H. Yadav, Z. Du, T. Joachims. Fair Learning-to-Rank from Implicit Feedback. SIGIR.2020

[16] J.Y. Wang , H.Y. Kao. RSOL: A Trust-Based Recommender System with an Opinion Leadership Measurement for Cold Start Users, Asia Information Retrieval Symposium. 2013.

[17] N.P. Andersen, Reducing Cold Start problem in the Wikipedia Recommender System. Master's thesis Academic thesis, 2011.

[18] P. Victor , MD Cock, C. Cornelis, et al. Getting cold start users connected in a recommender systems trust network. WORLD SCIENTIFIC, 2008.

[19] R. Salakhutdinov , A. Mnih . Probabilistic Matrix Factorization, NIPS, 2015.